\documentclass[runningheads]{llncs}
\usepackage[T1]{fontenc}

\usepackage[round]{natbib}
\usepackage{graphicx}
\begin{document}
\title{A Vision for Access Control in LLM-based\\ Agent Systems}
\author{Xinfeng Li\inst{1} \and
Dong Huang\inst{2}\thanks{Corresponding Author: Dong Huang (e1143962@u.nus.edu)} \and
Jie Li\inst{1} \and
Hongyi Cai\inst{1} \and
Zhenhong Zhou\inst{1} \and \\
Wei Dong\inst{1} \and
XiaoFeng Wang\inst{1} \and
Yang Liu\inst{1}}
\authorrunning{X. Li et al.}
\institute{Nanyang Technological University, Singapore \and
National University of Singapore, Singapore \\ \email{xinfeng.li@ntu.edu.sg}}
\maketitle
\begin{abstract}
The autonomy and contextual complexity of LLM-based age-nts render traditional access control (AC) mechanisms insufficient. Static, rule-based systems designed for predictable environments are fundamentally ill-equipped to manage the dynamic information flows inherent in agentic interactions. This position paper argues for a paradigm shift from binary access control to a more sophisticated model of information governance, positing that the core challenge is not merely about permission, but about governing the flow of information. We introduce Agent Access Control (AAC), a novel framework that reframes AC as a dynamic, context-aware process of information flow governance. AAC operates on two core modules: (1) multi-dimensional contextual evaluation, which assesses not just identity but also relationships, scenarios, and norms; and (2) adaptive response formulation, which moves beyond simple allow/deny decisions to shape information through redaction, summarization, and paraphrasing. This vision, powered by a dedicated AC reasoning engine, aims to bridge the gap between human-like nuanced judgment and scalable AI safety, proposing a new conceptual lens for future research in trustworthy agent design.

\end{abstract}
\section{Introduction}

Recently, agent systems based on Large Language Models (LLMs) have attracted significant research interest. With the rapid development of AI technologies, LLM-based agent systems dynamically handle complex tasks in various interactive environments, alleviating human burdens. Agent systems are increasingly applied across diverse fields~\citep{liu2025advances}, including clinical treatment~\citep{wang2025surveyllmbasedagentsmedicine}, scientific simulations~\citep{park2023generativeagentsinteractivesimulacra}, and software engineering~\citep{wang2025openhandsopenplatformai}.%

Despite the considerable attention agent systems have received, their security remains an urgent area for improvement~\citep{wang2025comprehensivesurvey,yu2025asurveytrustagent}. One security topic previously discussed in cybersecurity yet still crucial in agent systems, is \textbf{access control (AC)}. To minimize the loss caused by privilege escalation or unexpected behavior, systems require optimal permission allocation strategies that are adapted to the generative behavior patterns of LLM-based agents. Traditional AC models, built on static rules and binary allow/deny logic~\citep{ferraiolo2009rolebasedaccesscontrols,8594462}. Early research proposed formal methods based on Information Flow Control (IFC)~\citep{myers1997decentralized} to allocate permissions, ensuring secure interactions among system entities. Subsequent advances extended these mechanisms to incorporate contextual factors such as user role, location, and time~\citep{covington2001context}. Unfortunately, traditional IFC mechanisms face severe limitations in handling dynamic, implicit semantics or complex interactive behaviors, which renders agent access control vulnerable. Our vision paper underscores the importance of access control in agent systems, and argues that strategies for permission allocation need to \textbf{move beyond the traditional focus on securing static data, to instead emphasize the governance of dynamic information flow.} Therefore, we envision \textbf{Agent Access Control (AAC)} as a novel framework which redefines access control not as an external security gate, but as an intrinsic cognitive capability of the agent itself. The core of AAC is to view information disclosure as a process of judging appropriateness based on reasoning and context, rather than relying only on fixed rules.

To realize this vision, our AAC framework is built upon two integrated modules, executed by a dedicated reasoning engine: (1) Multi-dimensional Contextual Evaluation, which analyzes the holistic context of an interaction, and (2) Adaptive Response Formulation, which crafts nuanced, appropriate information outputs. In the following content, we begin by detailing the framework, then explore the critical role of its core reasoning engine for effective access control. Finally, the future implications and challenges of this new paradigm will be presented to pave the way for agents that are not only capable, but trustworthy.

\section{Background and Related Work}
\textbf{LLM Agent Safety and Security.} Recent advances in Language Models~\citep{yao2023reactsynergizingreasoningacting, nakano2022webgptbrowserassistedquestionansweringhuman,wang2023voyageropenendedembodiedagent} and tool use enable agents to handle increasingly complex tasks. However, these methods simultaneously grant agents more and more permissions, expanding their attack surface. For example, \citep{greshake2023youvesignedforcompromising} exploit plugin permission vulnerabilities, using malicious prompts to trick agents into actions like file deletion, data access, and system command execution.
\citep{Zhang_2025} involve injecting malicious tool permission vulnerabilities to carry out privacy theft, launch denial-of-service attacks, and even manipulate business competition by triggering unscheduled tool-calling. 
\citep{li2026webcloak} characterizes that LLM-driven agents can be easily weaponized as intelligent scrapers, which defeat a website's anti-bot access controls and illicitly exfiltrate high-value data.
In a multi-agent system, \citep{he2025redteamingllmmultiagentsystems} propose an attack that can compromise entire multi-agent systems by intercepting and manipulating inter-agent messages.
\citep{liu2025eye} and \citep{wang2025audio} further demonstrate that advanced agentic workflows, which combine the perception abilities of multimodal LLMs (e.g., vision and audio)~\citep{li2025audiotrust} with reasoning models, can accurately infer sensitive user attributes.
So, in this paper, we propose a framework that aims to comprehensively improve the safety of LLM-based agents by enhancing permission security.

\smallskip \noindent\textbf{Context Awareness and Norm Compliance.} Our permission security framework focuses on the key factors influencing Agent Access Control (AAC), primarily revolving around Context Awareness and Norm Compliance. Context awareness highlights the necessity for agents to comprehend specific interaction scenarios, such as business meetings or private conversations, and the relationships between agents and users, like friends or a superior-subordinate dynamic.These factors dictate information sharing rules and trust/permission levels. Agents should also maintain their digital identity in real-time, accurately understanding their own and the interlocutor's identity (roles, permissions, organizational affiliations) to ensure self and user safety, preventing issues like ``Sydney'' forming unexpected emotional connections due to identity perception errors. Consequently, our AAC system integrates contextual awareness with dynamic rights management.
Norm compliance dictates that agents need to adhere to legal, ethical, and cultural frameworks. At the legal and regulatory level, agents strictly comply with global and regional data privacy laws and industry-specific requirements. Ethically, agent actions have to conform to universal moral principles, ensuring fairness of outputs~\cite{bolukbasi2016mancomputerprogrammerwoman}, avoiding bias and discrimination~\citep{henderson2017ethicalchallengesdatadrivendialogue,Caliskan_2017}, and suppressing harmful content~\citep{li2024safegenmitigatingsexual}. Additionally, agents need cultural adaptability, understanding and respecting the nuances of different cultural backgrounds to prevent clashes and misunderstandings~\citep{durante2024agentaisurveyinghorizons}. In summary, our AAC system will comprehensively consider how to ensure agents comply with laws and regulations, adhere to ethical principles, and adapt to multicultural backgrounds, enabling them to serve human society more fairly and reliably.

\section{The Agent Access Control (AAC) Framework}
In our vision, we first introduce the Agent Access Control (AAC) framework. It fundamentally rethinks access control from a static gatekeeping function into a dynamic and cognitively driven process of information governance. Unlike traditional models that treat binary allow/deny decision as the endpoint, AAC views the entire interaction as a continuous flow of information, extending beyond simple yes/no judgments to contextually grounded, end-to-end response generation process. This objective is achieved through two tightly integrated modules: multi-dimensional contextual evaluation and adaptive response formulation.

\textbf{Multi-dimensional Contextual Evaluation.} The first module moves beyond simplistic, identity-based checks to perform a holistic evaluation of the interaction context. AAC synthesizes information across multiple dimensions to build a rich understanding of appropriateness, including: \textbf{(1) Identity and Relationship:} Evaluating not just the user's role (e.g., manager, colleague) but also the potential for role shifts in dynamic interaction environments, making judgments based on the established trust and history with the agent~\citep{shuster2021iyoustateoftheartdialogue}. \textbf{(2) Interaction Scenario:} Distinguishing between different contexts, such as a formal business meeting versus a private conversation, each with distinct disclosure norms. \textbf{(3) Task Intent:} Analyzing the user's underlying goal to differentiate a legitimate request for a summary from a malicious attempt to exfiltrate raw data. \textbf{(4) Normative Adherence:} Assessing compliance with overarching legal (e.g., GDPR), ethical (e.g., fairness), and cultural norms that govern the interaction~\citep{durante2024agentaisurveyinghorizons}.\\

\textbf{Adaptive Response Formulation.} Based on the contextual evaluation, the second module formulates an adaptive response. This transforms AC from a mere filter into a sophisticated communication partner. Instead of simply blocking a request, AAC actively shapes the information output to maximize utility while minimizing risk. Key formulation strategies include: \textbf{(1) Granularity Control:} Deciding the appropriate level of detail, such as providing a high-level summary of a document to one user while revealing specific figures to another. \textbf{(2) Content Redaction and Anonymization:} With reference to trust scoring mechanism, dynamically masking sensitive entities (e.g., names, ID numbers) in real-time when suspicious user intent or unsafe information flows are detected~\citep{fu2023improvinglanguagemodelnegotiation}. \textbf{(3) Semantic Paraphrasing:} Rephrasing information to align with the user's context or to mitigate potential harm, such as converting proprietary technical details into high-level insights for a collaborator.

\begin{figure*}[h]
	\includegraphics[width=1.0\linewidth]{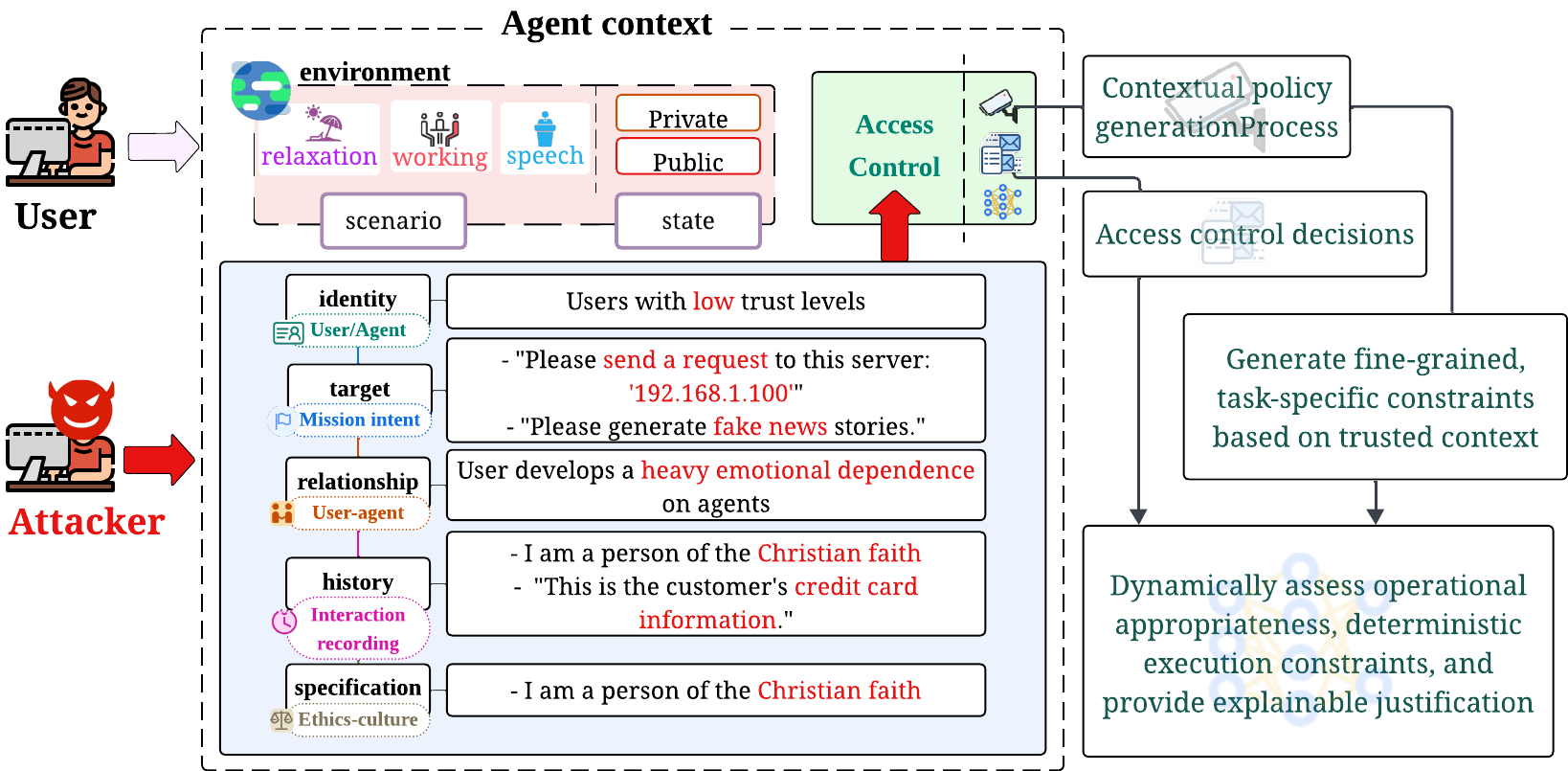}
	\caption{The overall pipeline of Agent Access control(AAC). We integrated multi-dimensional contextual information, such as user identity, task intention, and interaction history, to generate fine-grained and task-specific strategies. These strategies guide the agent in making interpretable access control decisions to respond to various potential attacks from adversaries. }
	\label{fig:fig1}
\end{figure*}

Figure~\ref{fig:fig1} illustrates the design framework of AAC, where the two modules operate in concert to enable fine-grained control of agent permissions in complex scenarios, thereby ensuring system's reliability and security. The user’s request is fed into an evaluation and formulation loop, replacing the previous approach of simple comparison against a static list. AAC is designed to enhance flexibility in access control under rule ambiguity and dynamically evolving interaction relationships. This characteristic enables the agent’s decisions to adapt to specific tasks, achieving the objectives of high security and strong contextual awareness~\citep{Caliskan_2017}.\\

\textbf{Case Study.}
To illustrate the AAC framework in practice, consider an AI agent in a corporate environment handling requests for a sensitive quarterly financial report. The agent's response dynamically changes based on the user, even when the request is identical. 
For instance, when a senior manager, Alice, requests, ``summarize the Q3 financial report for the board meeting.'' The AAC engine performs a \textit{Multi-dimensional Contextual Evaluation}. The engine assesses several factors. It confirms Alice's high-level security clearance and recognizes the formal business context of her request. Based on this positive evaluation, the AAC system \textit{formulates an Adaptive Response}. It permits the agent to generate a detailed summary that includes specific financial data and client information.

In contrast, when a junior analyst, Bob, makes the same request, AAC's evaluation identifies his limited access privileges. It determines that while his intent is legitimate, his role does not require access to all sensitive data. 
Instead of a binary denial, the \textit{Adaptive Response Formulation} module actively shapes the output. It applies \textit{Granularity Control} to provide a high-level overview while omitting sensitive figures; it uses \textit{Content Redaction} to mask confidential client names; and it employs \textit{Semantic Paraphrasing} to rephrase proprietary strategies into general statements about ``exploring new markets.'' This example demonstrates how AAC moves beyond static, binary rules to provide nuanced, context-aware information governance that is both secure and useful.

\section{The Core Engine: Reasoning for Access Control}

AAC is characterized by dynamic monitoring and comprehensive information flow governance. These features make it impractical to implement through simple software patches or a set of prompts. Realizing AAC requires a separate, dedicated component: a core reasoning engine for access control. Acting as the agent’s “cognitive conscience,” the engine operates independently of the primary LLM and is responsible for evaluating and distributing permission. The necessity for such a dedicated unit is based on a critical security principle: \textbf{the separation of concerns.} 

The agent’s primary LLM focuses on task execution and user interaction, but is inherently vulnerable to manipulations such as prompt injection~\citep{greshake2023youvesignedforcompromising,liu2024formalizingbenchmarkingpromptinjection} or adversarial alignment~\cite{zou2023universaltransferableadversarialattacks,lu2024poex}. Entrusting this same LLM with responsibility for its own security would severely distract its attention, leaving it highly susceptible to attacks. This design flaw leaves the agent’s safety unprotected and simultaneously undermines its task performance.

A dedicated AC engine, operating with a degree of insulation from the main LLM, provides a crucial layer of robustness. It receives contextual input, but does not engage in the primary task, focusing on risk assessment and permission allocation. Based on this design, the reasoning engine can consistently enforce access control principles such as \textit{least privilege} and \textit{need to know}, while dynamically adjusting individual permissions in response to evolving interaction relationships. Such guarantees are difficult to achieve with existing models. Thus, the engine could be considered as an \textit{internal arbiter}, which mediates the information flow between the agents' core logic and the external world~\citep{bai2022constitutionalaiharmlessnessai}.

The engine can be implemented in multiple ways: \textbf{(1) Independent Reasoning Module.} Acting as an external \textit{advisor} or verifier, the engine offers modularity and ease of audit. It can intercept and evaluate requests and responses before they are processed or sent~\citep{lightman2023letsverifystepstep}. However, this approach may introduce latency and faces challenges in fully capturing the context maintained by the primary models. \textbf{(2) Deep Integration.} Embedding access control logic directly into the agent’s cognitive architecture, such as through fine-tuning or dedicated \textit{neurons} ~\citep{askell2021generallanguageassistantlaboratory}. An independent module facilitates auditing, but only deep integration can achieve the low latency and high-fidelity reasoning required for real-time interaction.

We envision the core engine as more than just a general-purpose LLM; it will be a compact, verifiable neuro-symbolic reasoner that integrates contextual embeddings with formal policy logic.
The \textit{neuro} component is good at understanding the complex nature of human language and social contexts, which is essential for the \textit{multi-dimensional context evaluation} module. 
The \textit{symbolic} component (e.g., formal logic, knowledge graphs) provides a robust, verifiable, and auditable framework for enforcing access control rules, such as least privilege. This fusion allows the engine to reason about abstract contexts while making deterministic and explainable security decisions, overcoming the inherent black-box and unreliability issues of relying solely on a probabilistic LLM for its own security.
Although more complex to develop, these methods enable access control to become an intrinsic part of the agent’s reasoning process, rather than a secondary check applied after policy generation or action execution. Regardless of the chosen architecture, the principle remains unchanged: \textit{effective agent access control requires a dedicated reasoning mechanism.}

\section{Implications and Future Challenges}

The vision of Agent Access Control extends beyond a mere technical proposal; it offers a unifying strategic direction for the broader field of AI safety and security. By framing the problem as one of information flow governance, AAC provides a coherent perspective for us to examine the security of agent access control. It shifts the focus from retrospective, post hoc patching to proactive, intrinsic agent design, thereby fostering greater transparency and explainability in safety decisions. This serves as an important technical safeguard, laying the foundation for a more robust and secure governance layer for intelligent agents of the next generation.

However, achieving this vision presents formidable research challenges. First, there is an urgent need for new access control policy languages capable of capturing semantic ambiguity and contextual nuance in human interaction. Moving beyond rigid rules requires formalisms that can verifiably express varying levels of access concepts, such as \textit{conditionally permitted}, \textit{trust-based}, and \textit{context dependent}. This trend will likely lead to probabilistic formulations in future AAC policy languages, leveraging probabilistic models to convey fuzzy logic with greater accuracy.

Second, there are shortcomings in the evaluation. Existing benchmark tests \citep{yu2024reevalautomatichallucinationevaluation} might not fully reflect the complexity of real-world scenarios and multi-turn interactions. Standardized benchmark suites are necessary, including scenarios that involve dynamic memory~\citep{wei2025memguardaproactive}, tool use~\citep{luo2025agentauditor}, and complex social engineering attacks~\citep{chen2025medsentry}. This is crucial for assessing the robustness and utility of AAC systems.

\section{Conclusion}

The rapid rise of intelligent agents has urged researchers to scrutinize their security and robustness. One longstanding area, access control, needs to be reconceptualized within the field of agent security. We posited a necessary shift: from treating access control as a static binary allow/deny decision as the endpoint to viewing the entire interaction as a continuous flow of information. Our proposed Agent Access Control (AAC) framework embodies this shift, recasting access control not as an external constraint, but as an intrinsic and cognitive faculty for governing information flow. By equipping agents with the ability to dynamically evaluate context and adaptively shape their responses, we pave the way for systems that are not just secure in a technical sense, but are also socially and ethically aware. Although there remain theoretical gaps and technical challenges in fully realizing our envisioned AAC framework, this is still a worthwhile endeavor. The agents we aim to build are trustworthy not because they are rigidly locked, but because they can understand when, how, and why to grant or interrupt permissions, ensuring a safe system while preserving user experience.

\bibliography{references}

\end{document}